\documentstyle[12pt]{article}

\title{
{\vspace{-3cm} \normalsize \hfill
                        \parbox{33mm}{MS-TPI-93-10 \\
                                      hep-lat/9403014}  }\\[25mm]
Nondecoupling of Heavy Mirror-Fermion}

\author{ Lee~Lin \\
  Institut f\"ur Theoretische Physik I\\
  Universit\"at M\"unster\\
  Wilhelm-Klemm-Str.~9\\
  D-48149 M\"unster, FRG}

\date{Revised, January 21, 1994 }

\newcommand{\be}{\begin{equation}}
\newcommand{\ee}{\end{equation}}

\newcommand{\GP}{G_\psi}
\newcommand{\GC}{G_\chi}
\newcommand{\GRP}{G_{R\psi}}
\newcommand{\GRC}{G_{R\chi}}
\newcommand{\GRPTT}{G_{R\psi}^{(3)}}
\newcommand{\GRCTT}{G_{R\chi}^{(3)}}

\begin{document}
\maketitle

\begin{abstract} \normalsize
According to one-loop perturbation theory, fermions whose
masses are totally generated from Yukawa couplings
do not decouple in the heavy mass limit. We investigate this issue
nonperturbatively in the strong coupling regime of
the chiral $U(1)$ mirror-fermion model
in four dimensions. Our numerical results,
obtained on $6^3\cdot 16$, $6^3\cdot 24$
and $8^3\cdot 16$ lattices, indicate
nondecoupling of heavy fermion and mirror-fermion,
thus supporting the one-loop picture.
\end{abstract}

\section{ Introduction }

The Higgs and heavy fermion sectors in the Standard Model have
received a lot of attention from the lattice community recently.
The main difficulty in numerical studies of them is how to deal
with chiral fermions on the lattice. Different approaches were
proposed \cite{LAT}. The approach adopted in this letter
is the mirror-fermion method of Montvay \cite{IM}.

In the symmetric (PM) phase of the mirror-fermion models,
the fermion and its mirror partner are
always degenerate. Only in the broken (FM) phase where
the scalar field develops a nonzero vacuum expectation value (VEV)
can one manage to decouple the mirror-fermion.

One way to decouple the mirror-fermion is to decouple it like a
right-handed neutrino \cite{GPLW}. This is possible due to the shift
symmetry of the action at vanishing mirror-fermion Yukawa coupling.
Another possibility is to give the mirror-fermion a very heavy
mass (around the cut-off scale or almost as heavy as the
lowest doubler) and hopefully they will have vanishing effects
on the scalar field in this limit \cite{LMMW}.
This kind of decoupling is based on the decoupling theorem \cite{ACC}
saying that when the particle has a mass much higher than the physical
scale, it will have very small influence (e.g.
radiative corrections) on the ``physical world". However,
it can be easily seen at one-loop level that the contribution
of the particle to renormalized quantities of other particles will
not be suppressed by its huge mass if its mass is generated from the
Yukawa coupling, i.e. if its mass is generated through the mechanism
of spontaneous symmetry breaking (SSB). This is the
so-called phenomenon of nondecoupling.

Nondecoupling of heavy fermions in theories with SSB
has been discussed in several papers \cite{LC,DEC,AP}.
So far, all arguments are within (one-loop) perturbation theory
(except in \cite{AP} where large-N expansion is used).
Nondecoupling beyond one-loop is still not clear.
We decide to study this issue in a nonperturbative way
in our $U(1)$ scalar-fermion model with explicit mirror-fermions.
We still ignore the gauge field assuming that this approximation
will not change the picture qualitatively.
Hence, the scalar field is our ``physical world".

\section{ Action and Renormalized Quantities }

The Euclidean lattice action for the mirror-fermion model with
chiral $U(1)$ symmetry in four dimensions
in the usual normalization convention for numerical simulations is
$$
S = \sum_x \bigl\{ (\phi_x^+\phi_x) +
\lambda \left[ (\phi_x^+\phi_x) - 1\right]^2
-\kappa\sum_\mu (\phi^+_{x+\hat{\mu}}\phi_x)
+  \mu_{\psi\chi} \left[
(\overline{\chi}_x\psi_x) + (\overline{\psi}_x\chi_x) \right]
$$
$$
- K \sum_\mu \Bigl [
(\overline{\psi}_{x+\hat{\mu}} \gamma_\mu \psi_x) +
(\overline{\chi}_{x+\hat{\mu}} \gamma_\mu \chi_x)
+ r \left( (\overline{\chi}_{x+\hat{\mu}}\psi_x)
- (\overline{\chi}_x\psi_x)
+ (\overline{\psi}_{x+\hat{\mu}}\chi_x)
- (\overline{\psi}_x\chi_x)  \right) \Bigr ]
$$
\be \label{eq01}
+ \GP \left[ \overline{\psi}_x(\phi_{1x}-i\gamma_5\phi_{2x})\psi_x
         \right]
+ \GC \left[ \overline{\chi}_x(\phi_{1x}+i\gamma_5\phi_{2x})\chi_x
         \right]
\bigr\} \,.
\ee
Here $K$ is the fermion hopping parameter, $r$ the Wilson-parameter,
which will be fixed to $r=1$ in the numerical simulations. The complex
scalar field is $\phi_x\equiv\phi_{1x}+i\phi_{2x}$, and
$\Psi_x \equiv (\psi_x, \chi_x)$ stands for the mirror pair of fermion
fields (usually $\psi$ is the fermion and $\chi$ the
mirror-fermion), $\sum_\mu$ sums over eight directions
and $\gamma_{-\mu}=-\gamma_\mu$.
In this normalization the fermion--mirror-fermion mixing mass is
$\mu_{\psi\chi}=1-8rK$.

\vspace{0.3cm}
Renormalized quantities in the FM phase are defined as follows:

\noindent
For a given configuration of the scalar field, we measure
\be \label{eq02}
\phi\equiv{1\over L^3\, T}\sum_x\, \phi_x=|\phi|\, e^{i\alpha}\,\, .
\ee
The bare VEV is defined as
\be \label{eq03}
v\equiv\langle|\phi|\rangle\,\,\, .
\ee
For each given configuration, the longitudinal ($\phi_{Lx}$) and
transverse ($\phi_{Tx}$) scalar field components are defined as:
\be \label{eq04}
{\phi'}_x=e^{-i\alpha}\, \phi_x\equiv\, \phi_{Lx}\,
+\,i\,\phi_{Tx}\,\, .
\ee
At the same time, the fermion fields should be
rotated:
$$
{\psi'}_{Lx}=e^{-i\alpha}\psi_{Lx}\,\, , \,\,
{\psi'}_{Rx}=\psi_{Rx}\,\, ,
\,\, \bar{\psi'}_{Lx}=\bar\psi_{Lx}e^{i\alpha}\, \, ,
\,\, \bar{\psi'}_{Rx}=\bar\psi_{Rx}\, \, ,
$$
\be \label{eq05}
{\chi'}_{Rx}=e^{-i\alpha}\chi_{Rx}\,\, , \,\,
{\chi'}_{Lx}=\chi_{Lx}\,\, ,
\,\, \bar{\chi'}_{Rx}=\bar\chi_{Rx}e^{i\alpha}\,\,  ,
\,\, \bar{\chi'}_{Lx}=\bar\chi_{Lx}\,\, .
\ee

The reason why we make transformations on all fields
for each configuration is because there is actually no
SSB on a finite lattice. Numerical measurements of non-symmetric
quantities on a finite lattice will always end up with zero.
We therefore need to make the above rotations and define the bare
VEV as the length of $\phi$ in eq.(2). This quantity will
converge to the bare VEV of the scalar field
in the thermodynamic limit \cite{HJ}. In the same limit,
$\phi_{Lx}-v$ and $\phi_{Tx}$ will converge to
the Higgs and Goldstone fields.

In our numerical simulation in the FM phase, renormalized quantities
for the scalar field are defined by the timeslices of the
longitudinal and transverse correlations
$$
S_{Lt}\equiv
{1\over L^3}\,\sum_{\bf x}\bigl(\langle \phi_{L{\bf x}t}
\phi_{L{\bf 0}0}\rangle)_c=
{1\over L^3}\,\sum_{\bf x}\bigl(\langle \phi_{L{\bf x}t}
\phi_{L{\bf 0}0}\rangle-v^2\bigr)\,\, ,
$$
\be \label{eq06}
S_{Tt}\equiv
{1\over L^3}\,\sum_{\bf x}\bigl(\langle \phi_{T{\bf x}t}
\phi_{T{\bf 0}0}\rangle)_c=
{1\over L^3}\,\sum_{\bf x}\bigl(\langle \phi_{T{\bf x}t}
\phi_{T{\bf 0}0}\rangle)\,\, .
\ee
Here it is taken into account that $\langle\phi_{Lx}\rangle=v$ and
$\langle\phi_{Tx}\rangle=0$.

The physical Higgs boson mass $m_L$ is obtained from
\be \label{eq07}
S_{Lt}=a+b(e^{-m_L t}+e^{-m_L
(T-t)})\,\, .
\ee
The constant $a$ is due to finite size effects \cite{MW}.

The wave-function renormalization constants $Z_L$ and
$Z_T$ are defined as
$$
Z_L\equiv\Bigl(m_L^2+4\,{\rm sin}^2{\pi\over T}\Bigr)
\,\, \sum_x\,{\rm cos}\Bigl({2\pi\over T}t\Bigr)
\langle\phi_{Lx}\phi_{L0}\rangle_c\,\, ,
$$
\be \label{eq08}
Z_T\equiv\, 4\,{\rm sin}^2{\pi\over T}
\,\, \sum_x\,{\rm cos}\Bigl({2\pi\over T}t\Bigr)
\langle\phi_{Tx}\phi_{T0}\rangle\,\, .
\ee
Because of the existence of the massless Goldstone boson
in the FM phase, longitudinal
quantities, e.g. $Z_L$, cannot be defined at zero momentum.
Although the transverse mode does not suffer from
infrared problems, $Z_T$ cannot be defined at zero momentum
either because $\phi_{Tx}$ corresponds to a
massless particle which has no rest frame.

The renormalized VEV, denoted by $v_R$, is defined as
\be \label{eq09}
v_R\equiv\,{v\over \sqrt Z_T}\,\,\, .
\ee

Renormalized fermionic quantities can be defined from the 2-point
function $\tilde\Gamma_\Psi(p)$ of the transformed fermion
field $\Psi'=(\psi',\chi')$:
\be \label{eq10}
\tilde\Delta_\Psi(p)\equiv \sum_x\, e^{
-ip\cdot(y-x)}\,\langle\Psi_y'\bar\Psi_x'\rangle\,\, .
\ee
At momenta $p=(0,0,0,p_4)$ where $p_4$ is small, we have
\be \label{eq11}
\tilde\Delta_\Psi(p)\simeq
A-i\,\,{\rm sin}(p_4)\cdot\gamma_4\, B\,\,\, ,
\,\,\,\,\,\,\tilde\Gamma_\Psi(p)\equiv
\tilde\Delta_\Psi^{-1}(p)\simeq M+i\,\,
{\rm sin}(p_4)\cdot\gamma_4\, N\,\, ,
\ee
with
\be
M=\Bigl(A+{\rm sin}^2(p_4) BA^{-1}B\Bigr)^{-1}\,\, ,
\,\,\,\,\, N=A^{-1}BM=MBA^{-1}\, \, .
\ee

The fermion wave-function renormalization matrix
$Z_\Psi^{1/2}$ has to transform $N$ to the unit matrix:
\be \label{eq13}
(Z_\Psi^{1/2})^T\, N\, Z_\Psi^{1/2}\, =\, 1
\ee
where superscript $T$ means transpose of the matrix.
The renormalized fermion mass matrix is obtained as
\be \label{eq14}
   M_R\equiv (Z_\Psi^{1/2})^T\, M\, Z_\Psi^{1/2} = \left(
         \begin{array}{cc}
          \mu_{R\psi}& \mu_R\\
          \mu_R      & \mu_{R\chi}
         \end{array}
         \right)
\,\, ,\,\,\,
G_{R\psi ,\chi}\equiv{\mu_{R\psi ,\chi}\over v_R}\,\, .
\ee

In general, $\mu_R\ne 0$, we therefore need to diagonalize
the renormalized mass matrix $M_R$ to obtain the physical
fermion states , which will be mixtures of the original
$\psi$ and $\chi$ states with the mixing angle $\alpha_R$
defined later. If the diagonalized mass matrix is denoted
by $M_{phys}$, we have
$$
M_{phys}= \left(
         \begin{array}{cc}
          \mu_{1R}& 0\\
          0       & \mu_{2R}
         \end{array}
         \right)
\,\, ,
$$
$$
\mu_{1R}=\mu_{R\psi}\,{\rm cos}^2(\alpha_R)+\mu_R\,{\rm sin}
(2\alpha_R)+\mu_{R\chi}\,{\rm sin}^2(\alpha_R)\equiv
 G_{1R}\, v_R\,\, ,
$$
$$
\mu_{2R}=\mu_{R\chi}\,{\rm cos}^2(\alpha_R)-\mu_R\,{\rm sin}
(2\alpha_R)+\mu_{R\psi}\,{\rm sin}^2(\alpha_R)\equiv
 G_{2R}\, v_R\,\, ,
$$
\be \label{eq15}
2\,\alpha_R={\rm tan}^{-1}\Bigl
({2\mu_R\over \mu_{R\psi}-\mu_{R\chi}}\Bigr)
\,\, .
\ee
Our convention is always that after diagonalization of $M_R$,
the second physical fermion state is the heavier one, i.e.
$|\mu_{2R}|>|\mu_{1R}|$.
The sign of the fermion mass is insignificant, because it
can be changed by an appropriate $\gamma_5$-transformation \cite{F}.

The 3-point renormalized Yukawa couplings coupled to the transverse
scalar field $\phi_T$ (i.e. the Goldstone
boson)\footnote{ Here one can think of
the Goldstone field as the longitudinal component of the gauge
field once the system is gauged.} are defined by
\be \label{eq16}
\left(
\begin{array}{cc}
 i\, \gamma_5\, \GRPTT & 0\\
  0      &-i\,\gamma_5\, \GRCTT
\end{array}
\right) \delta_{k,-p+q}
= {{\hat k}^2_4\over\sqrt{Z_T}}\,\tilde{\Gamma}_R(p_4)\, Z_\Psi^{-1/2}
\,G^{(c)}\, (Z_\Psi^{-1/2})^T\, \tilde{\Gamma}_R(q_4)\, \ ,
\ee
where $k_4$, $p_4$, $q_4$ are
the $4$th components of the momenta of Goldstone boson, fermion and
anti-fermion, respectively, and $\hat k_4^2$ is $4\,\rm{sin}^2(k_4/2)$.
We have set the spatial components of all momenta to zero.
The appearance of the Kronecker-delta above is due to
energy-momentum conservation where
$$
G^{(c)} = \frac{1}{L^3T}\,
\sum_{x,y,z}\,e^{-ik_4x_4}\,e^{-ip_4y_4}\,
e^{iq_4z_4}\Bigl<\phi_{Tx}\Psi_y'\bar\Psi_z'\Bigr >_c
$$
is the connected part of the $\phi_T$-$\Psi'$-$\bar\Psi'$
3-point Green's function. Since $\langle\phi_{Tx}\rangle=0$,
the above connected Green's function is equal to the disconnected one.
In our simulations on $L^3 \cdot T$ lattices we choose
$$
k_4 = {2\pi\over T}\,\, ,\,\,
p_4 = -{\pi\over T}\,\, ,\,\, q_4={\pi\over T}
\,\, .
$$
 After carrying out all the matrix multiplications, we
 get $\GRPTT$ and $\GRCTT$.
 The expressions are too voluminous to be displayed here.

If $\mu_R\ne 0$, one more diagonalization
is necessary to obtain the 3-point renormalized Yukawa
couplings for the physical fermion states, denoted by
$G_{1R}$ and $G_{2R}$ respectively. We have
\be \label{eq17}
G_{1R}^{(3)}=\GRPTT{\rm cos}^2(\alpha_R)-
\GRCTT{\rm sin}^2(\alpha_R)\,\,\, ,\,\,
G_{2R}^{(3)}=\GRCTT{\rm cos}^2(\alpha_R)-
\GRPTT{\rm sin}^2(\alpha_R)\,\, \, .
\ee
We then define the following ratios:
\be \label{eq18}
R_\psi\equiv{\GRPTT\over G_{R\psi}}\,\, ,\,\,
R_1   \equiv{G_{1R}^{(3)}\over G_{1R}}\,\,\,\, ;\,\,\,\,
R_\chi\equiv{\GRCTT\over G_{R\chi}}\,\, ,\,\,
R_2   \equiv{G_{2R}^{(3)}\over G_{2R}}\,\,\,\, .
\ee
Note that renormalized
fermionic quantities defined from eq.(\ref{eq11})
to eq.(\ref{eq18}) are valid at the zero corner of the Brillouin zone,
i.e. they represent properties of the two physical fermions. One can
actually define corresponding quantities for fermions at other corners
in the Brillouin zone. For instance, to obtain quantities for the
fermion doublers at $p=(e_1\pi,e_2\pi,e_3\pi,e_4\pi)$; $e_{1,2,3,4}
=0,1$, one simply makes the following transformation on the
rotated fermion field
\be
{\Psi'}_x\rightarrow{\Psi'}^t_x\, =(-1)^{e_1 x_1+e_2 x_2+e_3 x_3
+e_4 x_4}\,{\Psi'}_x\,\,\, .
\ee
Then fermionic renormalized quantities for the two doublers
sitting at $p$ will be obtained from eq.(\ref{eq11}) to eq.(\ref{eq18})
by substituting the transformed fermion field for the original one.
Renormalized fermionic quantities for the doublers will have the
same notations as their corresponding quantities for the
two physical fermions except that we attach a bracket in which we
denote where in the Brillouin zone they are sitting. For example,
ratios $R_1$ and $R_2$ for the doublers at one-$\pi$ corner are
denoted as $R_1(1\pi)$ and $R_2(1\pi)$ respectively.

\vspace{0.3cm}
As we mentioned earlier, one
possible way of dealing with the mirror-fermion is to give it a very
heavy mass and hopefully it will have vanishing effects on the
scalar field in this limit \cite{LMMW}. This can be easily seen
at one-loop level where the scalar propagator receives
radiative correction from the fermion loop whose contribution
is proportional to the square of the
ratio of the two Yukawa couplings (e.g.
$R_\psi^2$ in (\ref{eq18})). In the heavy fermion
mass limit, this ratio will be suppressed by the
factor of fermion mass. However, in a theory with SSB,
the Yukawa coupling itself is proportional to the fermion
mass \cite{LC}.
In this case, the ratio say, $R_\psi$ will be a constant
and the radiative correction will not be suppressed by the heavy fermion
mass. Instead, it will be proportional to $1/v_R^2$,
hence the heavy fermion will not decouple from the real world.
It is necessary to go beyond
one-loop and investigate this issue in a completely
nonperturbative way, i.e. on the lattice.

Since $\mu_R\ne 0$ in general,
the relevant ratio to measure for the study of
decoupling of heavy mirror-fermion is $R_2$, which is very
close to $R_\chi$ when $\mu_R\simeq 0$.
According to one-loop calculation, 3-point Yukawa couplings defined
in (\ref{eq16}) coincide with $G_R$'s defined by the mass-to-VEV
ratio in (\ref{eq14}).
When the Yukawa couplings are strong, they may differ from
each other and $G_R$'s in (\ref{eq14}) cease to be a good
definition for the renormalized Yukawa couplings. There, the
better definition should be the ones in (\ref{eq16}).
What is interesting to see is whether the two Yukawa couplings
still coincide with each other beyond one-loop level.
If this is the case in the strong-coupling regime such that
$R_2=1.0$ within the error, then we take it as
the nonperturbative signal for nondecoupling.

At tree level, it is
clear that $\mu_R$ for the doublers will have contributions from
the Wilson term. Therefore we expect
that masses for the ``physical" (i.e. diagonalized)
doublers will not be totally generated from Yukawa interactions,
and decoupling theorem should apply to all of them.
Corresponding ratios $R_1$ and $R_2$ are also measured to see
whether doublers are decoupled as expected.

\section{ Numerical Simulation }

The Monte Carlo simulations were performed by the unbiased
Hybrid Monte Carlo method \cite{HBMC}.
Therefore, the fermions have to be doubled by taking the
adjoint of the fermion matrix for the second species.
(The fermionic part in (\ref{eq01}) is given for a single mirror
pair of fermions.) The number of leapfrog steps per molecular
trajectory was chosen randomly between 3 and 10. The step size was
tuned to maintain an acceptance rate around $75\%$. The necessary
inversions of the fermion matrix were done by the conjugate
gradient iteration, until the residuum was smaller than
$10^{-12}$ times the length square of the input vector on the
$6^3\cdot 16$ and $6^3\cdot 24$ lattices. On the $8^3\cdot 16$
lattice, a precision of $10^{-13}$ was necessary.

In the simulations, we set
$\lambda=\infty$ and $\GC=-0.1$, $-0.3$, $-0.6$ and $-1.0$, while
$\GP$ is kept at 0.1. The mass parameters $\kappa$ and $K$ are tuned
such that the mixing mass in the FM phase is zero or small, and
the renormalized fermion mass $\mu_{1R}$ is around 120
$GeV$ in physical units, mimicking the top quark.
(The scale is set by $v_R=246$ $GeV$.)
We let $\GP$ and $\GC$ have opposite signs because this
combination is preferred by the condition of reflection
positivity \cite{LMMW}. The mixing mass $\mu_R$
is tuned to be small in order that masses of
the physical fermions are basically completely generated from
SSB, which is the case we interested in.
The most difficult task in this numerical simulation is the
tuning of $K$ to have $\mu_R\simeq 0.0$.
(Our criterion for $\mu_R\simeq 0.0$ is that
$\mu_{R\psi}$, $\mu_{R\chi}$,
$\GRPTT$ and $\GRCTT$ do not differ from
$\mu_{1R}$, $\mu_{2R}$, $G_{1R}^{(3)}$ and
$G_{2R}^{(3)}$ respectively by more than $1\%$. In this case, the
second physical fermion state is basically the mirror-fermion.)
When $\GP\cdot\GC=0$, $K_c$ (the value of $K$ at which $\mu_R=0.0$)
is at $1/8$. When the bare Yukawa couplings
are small and $\GP\cdot\GC<0$, $K_c$ will be slightly larger
than $1/8$ according to one-loop bare perturbation
theory \cite{F}. However, we find that
even when $-\GC\ge 0.6$, $K_c$ still keeps growing as
is qualitatively predicted by one-loop result.

Most of the runs were done on the $6^3\cdot 16$ lattice.
Since it is not a very large lattice, we need to
check the finite size effect. This can be estimated by
one-loop renormalized perturbation theory on the lattice.
Formulae of one-loop bare perturbation theory on the lattice
were already presented in \cite{LMMW}.
Starting from those equations, one can obtain
one-loop renormalized perturbation theory on the lattice
where counterterms are introduced on the infinite lattice to
eliminate ultraviolet divergences. For example, the bare
fermion mass is given by eq.(74) in \cite{LMMW}.
By assuming that $Z_\psi=1/2K$ and stays constant as we change
the lattice size (which is a good approximation to our
numerical data), we obtain the finite size shift of the
renormalized fermion mass $\Delta \mu_{1R}\equiv
\mu_{1R}(\infty,\infty)-\mu_{1R}(L,T)$ at $\mu_R=0$ as
\be \label{eq20}
\Delta\mu_{1R}=\mu_{1R}\,\GRP^2\,\,\,\Delta\int_q
\Bigl[{1\over \hat{q}^2+m_L^2}-{1\over\hat{q}^2}\Bigr]
\,\tilde D(q)^{-1}\,(\bar{q}^2+\mu_{2R}^2)\,\,
\ee
where
$$
\Delta\int_q\equiv{1\over L^3\cdot T}\sum_q\, -
\int{d^4 q\over(2\pi)^4}\,\, ,\,\,
\hat{q}^2=4\sum_\mu{\rm sin}^2({q_\mu\over 2})\,\, ,
\,\, \bar{q}^2=\sum_\mu{\rm sin}^2(q_\mu)\,\, ,
$$
$$
\tilde D(q)=\bigl(\bar{q}^2-\mu_{1R}\,\mu_{2R}\bigr)^2+\bar{q}^2
(\mu_{1R}+\mu_{2R})^2\,\, .
$$
Notice that all quantities appearing on the right-hand side of
eq.(\ref{eq20}) are those in the infinite volume limit.
Similarly, one can derive one-loop results of finite volume
shifts for other renormalized quantities. For instance,
one should start from eq.(68) in \cite{LMMW} to derive
$\Delta m_L$. Those formulae are too voluminous
to be displayed here.

According to our previous experience, scalar quantities
usually suffer from finite size effects the most
due to small fermion mass $\mu_{1R}$.
{}From one-loop results, we do find that at $\GC=-1.0$
the effect amounts up to $10\%$ for $m_L$, while
$\GRC$ and $\GRCTT$ are smaller by about $5\%$
in the infinite volume limit.

Also, it is important to have some idea about the
finite cutoff effect since sometimes the cutoff is low.
The way we estimate finite cutoff effect is to compare
one-loop $\beta$-functions in the continuum and on the
lattice. One-loop $\beta$-functions in the continuum are given
by eqs.(83), (84) in \cite{F}.
Starting from eqs.(74) to (79) in \cite{LMMW}
one can obtain one-loop formulae for $G_R$'s
on the lattice. By differentiation with respect to the
energy scale, we will obtain their one-loop
lattice $\beta$-functions. The derivation and final results are
very lengthy and will not be stated here. Those one-loop results
show that at $\GC=-1.0$, as we go to smaller
mass scale say, from $m_L=1.3$ to 1.0, both $\GRC$ and
$\GRCTT$ decrease. But the decrease of them is faster
for the continuum $\beta$-function. We estimate that at
$m_L=1.0$, the cutoff dependence of both $\GRC$ and $\GRCTT$
is around $10\%$. Of course, this is just a crude estimate.
We also see that according to one-loop calculations, finite
volume and cutoff effects for $\GRC$ are very close to
those for $\GRCTT$ (with the same sign). This indicates
that the ratio $R_2$ should be quite stable against
those effects at one-loop level.

It is always better to check the finite volume and cutoff
effects numerically at the same time,
since one-loop formulae break down in the strong-coupling regime.
Due to limitations on CPU time, the numerical check
was carried out only at $\GC=-1.0$, which is the
most relevant point, on $6^3\cdot 24$
and $8^3\cdot 16$ lattices.
There we had to go to the negative-$\kappa$ region
in order to increase the cutoff. Although
reflection positivity cannot be proven when
$\kappa < 0$ \cite{LMMW}, data obtained there did not show anything
abnormal. We simply assume that reflection positivity
is still preserved at least in the region with
small negative $\kappa$.

\section{ Conclusions and Discussions }

Our numerical data are presented in tables 1, 2 and 3.
{}From data at points d, D, e, and E,
finite size effect for $m_L$ at $\kappa=-0.01$, $\GC=-1.0$ is quite
strong as expected. It is even stronger than predicted
by one-loop formula. The relevant quantities
$\GRC$ and $\GRCTT$ show smaller finite volume shifts, and
$R_2$ is quite stable as we increase the lattice size.
Finite cutoff effect in $\GRCTT$ is smaller
than predicted by one-loop result. But fluctuations at
point E are quite large.

Readers should note that at point cc, the mixing mass $\mu_R\ne 0$,
which means the mass of the second physical fermion has
contribution from the ordinary (off-diagonal) mass term.
Therefore, it is not surprising that $R_2 < 1.0$ there, showing
``partial" decoupling.
For all other points, parameter $K$ is tuned such that $\mu_R$ is
quite close to zero.
There the ratio $R_2$ is always equal to 1.0 within the error
up to $\GC=-1.0$ which, in our model, is a strong coupling case.
Hence, it is obvious from our data
that the heavy mirror-fermion does have its renormalized
Yukawa coupling proportional to the mass and
will not decouple from the transverse scalar field.
The one-loop picture survives the strong Yukawa coupling
limit. The idea of decoupling the mirror partners
by giving them large masses does not work.
Since the action in (\ref{eq01}) has
$\GP\rightleftharpoons\GC$, $\gamma_5\rightleftharpoons
 -\gamma_5$ symmetry, it is obvious that the heavy fermion itself will
not decouple either. It is also obvious that doublers at
one-$\pi$ corner do decouple. Fluctuations of data of higher doublers
are too large. But we believe that all doublers decouple like the
lowest ones do.

We think that these conclusions also apply to
the $SU(2)$ version of our mirror-fermion model
because its qualitative behaviour appears to be similar to
that of the $U(1)$ model \cite{FLMMPTW}.
We conjecture that in other scalar-fermion
models, heavy fermions do not decouple either.

\vspace{0.3cm}
A direct measurement of $\tilde S$ defined
defined in eq.(2) in \cite{AP} will be
overwhelmed by fluctuations for our models. Reasonable statistics
can be reached if we study a scalar-fermion
model with fewer internal degrees of freedom like the model with
staggered fermion coupled to one-component scalar field.

Once the chiral gauge field is put in, the shift symmetry at
$\GC=\mu=0$ in our models is broken. This means that the
mirror-fermion can no longer be decoupled as right-handed
neutrino in the presence of gauge field. Some other means
for decoupling should be used. We conjecture that
the mirror-fermion will remain coupled to the scalar field
in the heavy mass limit when the system is gauged. The more
serious problem there is actually the renormalized gauge
coupling. In order to preserve the chiral gauge symmetry in the
action, both $\psi_L$ and $\chi_R$ have to be gauged. It is
possible that in the FM phase
once $\psi_L$ is coupled to the gauge field
(i.e.: the real world), $\chi_R$ is also coupled at the same
time as is shown at one-loop level. If this is true
nonperturbatively, then mirror-fermion cannot be decoupled
eventually and the gauged mirror-fermion models have trouble in
reproducing the minimal Standard Model in the continuum limit.
This issue will be explored in details in the future.

\vspace{0.5cm}
I would like to thank I.\ Montvay and G.\ M\"unster
for useful discussions.

\newpage

\newpage
%
\begin{table}[tb]
\caption{  \label{t1} Data on renormalized VEV and masses are presented.
Points with lower-case letters are obtained on $6^3\cdot 16$
lattice while point D is on $6^3\cdot 24$ lattice,
point E on $8^3\cdot 16$ lattice. Each point has about
4000 trajectories for equilibration, and around 10000 to 12000
for measurements.
}
\begin{center}
\begin{tabular}
{|c|r@{.}l|r@{.}l|r@{.}l|r@{.}l|r@{.}l|r@{.}l|r@{.}l|r@{.}l|}
\hline
&\multicolumn{2}{c|}{$\GC$} & \multicolumn{2}{c|}{$\kappa$}
&\multicolumn{2}{c|}{$K$}
&\multicolumn{2}{c|}{$v_R$}&\multicolumn{2}{c|}{$m_L$}
&\multicolumn{2}{c|}{$\mu_R$}
&\multicolumn{2}{c|}{$\mu_{1R}$} & \multicolumn{2}{c|}{$|\mu_{2R}|$}
\\
\hline
 a & -0&1 &  0&160 & 0&1256 & 0&330(11) & 1&06(4)
   & 0&033(6)&  0&2008(4) & 0&2013(4) \\
\hline
 b & -0&3 &  0&126 & 0&128  & 0&205(11) & 0&77(6)
   & 0&041(10)&  0&118(2)  & 0&343(6)  \\
\hline
 c & -0&6 &  0&083 & 0&131  & 0&28(2)   & 1&01(6)
   & 0&040(12)&  0&1426(9) & 0&804(8)  \\
\hline
 cc& -0&6 &  0&083 & 0&1256 & 0&214(17) & 0&74(6)
   & 0&226(10)&  0&183(1) & 0&729(18)  \\
\hline
 d & -1&0 &  0&0   & 0&1345 & 0&248(3)  & 1&31(8)
   & 0&052(20)&  0&143(1)  &  1&29(3) \\
\hline
 D & -1&0 &  0&0   & 0&1345 & 0&240(8)  & 1&21(7)
   & 0&037(37)&  0&144(8)  &  1&22(7) \\
\hline
 e & -1&0 &  -0&01 & 0&1345 & 0&229(31)  & 1&35(10)
   & 0&070(31)  &  0&124(2) &  1&18(3) \\
\hline
 E & -1&0 &  -0&01 & 0&1345 & 0&231(20)  & 1&03(17)
   & 0&066(20)  &  0&118(2) &  1&026(42) \\
\hline
\end{tabular}
\end{center}
\end{table}
%
%
\begin{table}[tb]
\caption{  \label{t2} Data on renormalized Yukawa couplings
for the fermion and mirror-fermion. Labels are the same as in table 1.
}
\begin{center}
\begin{tabular}
{|c|r@{.}l|r@{.}l|r@{.}l|r@{.}l|r@{.}l|r@{.}l|}
\hline
&\multicolumn{2}{c|}{$\GRP$} & \multicolumn{2}{c|}{$\GRPTT$}
&\multicolumn{2}{c|}{$\GRC$} & \multicolumn{2}{c|}{$\GRCTT$}
&\multicolumn{2}{c|}{$G_{2R}^{(3)}$}
&\multicolumn{2}{c|}{$R_2$} \\
\hline
 a & 0&60(2) & 0&60(3) & -0&60(2) & -0&59(3)  & -0&59(3) & 0&99(4)\\
\hline
 b & 0&56(3) & 0&56(4) & -1&66(9) & -1&70(9)  & -1&70(9)& 1&02(3)\\
\hline
 c & 0&50(4) & 0&48(4) & -2&87(21)& -2&89(22) & -2&89(22)& 1&01(3)\\
\hline
 cc& 0&57(3) & 0&59(5) & -3&13(19)& -3&07(19) & -2&90(18)& 0&85(4)\\
\hline
 d & 0&57(5) & 0&63(6) & -5&19(49)& -5&09(48) & -5&08(48)& 0&98(6)\\
\hline
 D & 0&60(9) & 0&57(11)& -5&10(27)& -5&03(32) & -5&03(31)& 0&99(6)\\
\hline
 e & 0&52(4) & 0&53(5) & -4&92(33)& -4&67(34) & -4&66(34)& 0&95(6)\\
\hline
 E & 0&51(4) & 0&525(54)&-4&44(41)& -4&61(61) & -4&60(62)& 1&04(10)\\
\hline
\end{tabular}
\end{center}
\end{table}
\nopagebreak
%
\begin{table}[tb]
\caption{  \label{t3} Data on masses and couplings for the
two fermion doublers at one-$\pi$ corner.
Labels are the same as in table 1.
}
\begin{center}
\begin{tabular}
{|c|r@{.}l|r@{.}l|r@{.}l|r@{.}l|r@{.}l|r@{.}l|}
\hline
&\multicolumn{2}{c|}{$\mu_{1R}(1\pi)$}
&\multicolumn{2}{c|}{$|\mu_{2R}(1\pi)|$}
&\multicolumn{2}{c|}{$G_{1R}^{(3)}(1\pi)$}
&\multicolumn{2}{c|}{$G_{2R}^{(3)}(1\pi)$}
&\multicolumn{2}{c|}{$R_1(1\pi)$}
&\multicolumn{2}{c|}{$R_2(1\pi)$}
\\
\hline
 a & 2&00(1) & 2&00(1) & 0&63(11) & -0&59(11) & 0&10(3) &  0&10(3)\\
\hline
 b & 1&90()1 & 2&09(1) & 1&19(14) & -1&32(15) & 0&13(3)&  0&13(2)\\
\hline
 c & 1&65(2) & 2&30(3) & 1&13(16) & -1&65(19) & 0&19(2)&  0&20(2)\\
\hline
 cc& 1&87(3) & 2&39(3) & 1&93(17) & -2&36(21) & 0&22(2)&  0&21(2)\\
\hline
 d & 1&44(3) & 2&54(6) & 2&13(20) & -3&68(35) & 0&37(4) &  0&36(4)\\
\hline
 D & 1&45(18) & 2&78(22) & 1&96(22) & -3&82(57) & 0&32(4) & 0&33(4)\\
\hline
 e & 1&48(3) & 2&41(6) & 2&04(21) & -3&31(30) & 0&32(4) & 0&31(3) \\
\hline
 E & 1&50(4) & 2&40(10) & 1&85(21) & -3&21(53) & 0&28(3) & 0&31(5)\\
\hline
\end{tabular}
\end{center}
\end{table}

\begin{thebibliography}{99}
%
\bibitem{LAT}
I.\ Montvay, Nucl.\ Phys.\ B(Proc.\ Suppl.)26 (1992) 57;\\
D.N.\ Petcher, Nucl.\ Phys.\ B(Proc.\ Suppl.)30 (1993) 50.
%
\bibitem{IM}
I.\ Montvay,
Phys.\ Lett.\ B199 (1987) 89;
Nucl.\ Phys.\ B (Proc.\ Suppl.)4 (1988) 443.
%
\bibitem{GPLW}
M.F.L.\ Golterman, D.N.\ Petcher, Phys.\ Lett.\
225B (1989) 159;\\
L.\ Lin, H.\ Wittig, Z.\ Phys.\ C54 (1992) 331.
%
\bibitem{LMMW}
L.\ Lin, I.\ Montvay, G.\ M\"unster, H.\ Wittig,
Nucl.\ Phys.\ B355 (1991) 511.
%
\bibitem{ACC}
T.\ Appelquist, J.\ Carazzone, Phys.\ Rev.\ 11 (1975) 2856;\\
J.C.\ Collins, Renormalization
(Cambridge Univ.\ Press, Cambridge, 1984), p.223.
%
\bibitem{LC}
L.-F.\ Li, T.P.\ Cheng, in: The Vancouver Meeting,
Particles and Fields '91,
eds.\ D.\ Axen et.\ al., (World Scientific, Singapore, 1992) p.801.
%
\bibitem{DEC}
E.\ D'Hoker, E.\ Fahri, Nucl.\ Phys.\ B248 (1984) 59, 77;\\
D.C.\ Kennedy, P.\ Langacker, Phys.\ Rev.\ Lett.\ 65
(1990) 2967;\\
D.C.\ Kennedy, Mod.\ Phys.\ Lett.\ A6 (1991) 1459;\\
M.\ Dugan, L.\ Randall, Nucl.\ Phys.\ B382 (1992) 419;\\
T.\ Banks, Phys.\ Lett.\ B272 (1991) 75.
%
\bibitem{AP}
K.\ Aoki, S.\ Peris, Phys.\ Rev.\ Lett.\ 70 (1993) 1743.
%
\bibitem{HJ}
A.\ Hasenfratz et.\ al., Nucl.\ Phys.\ B317 (1989) 81;\\
K.\ Jansen, Nucl.\ Phys.\ B(Proc.\ Suppl.)4 (1988) 422.
%
\bibitem{MW}
I.\ Montvay, P.\ Weisz, Nucl.\ Phys.\ B290 (1987) 327.
%
\bibitem{F}
K.\ Farakos et.\ al., Nucl.\ Phys.\ B350 (1991) 474.
%
\bibitem{HBMC}
S.\ Duane, A.D.\ Kennedy, B.J.\ Pendleton, D.\ Roweth,
Phys.\ Lett.\ B195 (1987) 216.
%
\bibitem{FLMMPTW}
C.\ Frick, L.\ Lin, I.\ Montvay, G.\ M\"unster, M.\ Plagge,
T.\ Trappenberg, H.\ Wittig,
Nucl.\ Phys.\ B397 (1993) 431;
Nucl.\ Phys.\ B(Proc.\ Suppl.)30 (1993) 647;\\
L.\ Lin, I.\ Montvay, G.\ M\"unster, M.\ Plagge,
H.\ Wittig,
Phys.\ Lett.\ B317 (1993) 143.
%
\end{thebibliography}
\end{document}